\documentclass[aps,prl,preprint,superscriptaddress,floatfix]{revtex4-1}
\usepackage{graphicx}

\usepackage{amsmath}
\usepackage{amssymb}
\usepackage{hyperref}
\usepackage{bm}

\begin{document}
\title{Broadband achromatic anomalous mirror in near-IR and visible frequency range}%

\author{Andrei Nemilentsau}
\affiliation{Department of Electrical \& Computer Engineering, University of Minnesota, Minneapolis, MN 55455, USA}
\email{anemilen@umn.edu}
\author{Tony Low}
\email{tlow@umn.edu}
\affiliation{Department of Electrical \& Computer Engineering, University of Minnesota, Minneapolis, MN 55455, USA}


\begin{abstract}
The anomalous achromatic mirror operating in near-IR and visible frequency range was designed using an array of metal-insulator-metal (MIM) resonators. An incident wave interacting with MIM resonator experiences phase shift that is equal to the optical path travelled by the gap plasmon, excited by the wave. The phase gradient along the mirror surface is created through the difference in plasmons optical paths in resonators of different lengths. In the frequency region well below the plasma frequency of the metal, the phase gradient is a linear function of frequency, and thus the mirror operates in achromatic regime, i.e. reflection angle does not depend on the radiation frequency. Using silver-air-silver resonators, we predicted that the mirror can steer normally incident beam to angles as large as 40$^{\circ}$ with high radiation efficiency (exceeding 98 $\%$) and small Joule losses (below 10 $\%$).
\end{abstract}

\maketitle

\section{Introduction}
Recent progress in the design of gradient metasurfaces allows for the unprecedented control of characteristics of propagating light beams \cite{Kildishev13,Estakhri16}. This includes anomalous reflection and refraction, that defies conventional Snell's law\cite{Yu11,Larouche12,Ni12,Sun12,Pors13,Lin14,Nikitin14,Li15,Decker15}, efficient manipulation of the polarization state of electromagnetic waves \cite{Kang12,Pfeiffer13a,Yang14,Wu14,Arbabi15}, and polarization dependent steering of light beams \cite{Yin13,Pors13a,Farahani13,Shaltout15}. Electrically tunable graphene \cite{Fallahi12,Carrasco13,Carrasco15} and conducting oxide \cite{Huang16} metasurfaces, as well as thermally tunable dielectric metasurfaces \cite{Sautter15} were recently demonstrated. A number of applications utilizing these nontrivial properties of  metasurfaces were proposed, such as ultrathin flat lenses \cite{Aieta12,Chen12,Monticone13,Arbabi15a}, perfect absorbers \cite{Yao14,Akselrod15}, invisibility cloaks\cite{Ni15}, polarization detectors \cite{Pors15} and waveplates \cite{Zhao11,Nanfang12}, vortex and Bessel beam generators \cite{Cai12,Huang12,Genevet12,Pfeiffer13}, and holograms \cite{Huang13,Ni13,Chen14,Zheng15}.

Despite all the success achieved in engineering of gradient metasurfaces with non-trivial optical properties, their performance is usually restricted to a narrow frequency range and suffers from chromatic aberrations, such as frequency dependent angle of reflection in the case of anomalous mirror. One can argue that the limitation on the operation frequency range is inherent to the metasurface design. Indeed, basic element of the typical metasurface is a sub-wavelength scatterer, dielectric or metallic, upon interacting with which electromangetic wave experiences drastic phase change. The magnitude of the phase change depends on the detuning between frequency of electromagnetic wave and resonance frequency of the scatterer, and thus spatial gradient of a phase change can be created by placing resonators of different shapes and sizes along the metasurface. It is this phase gradient that is responsible for all the unique properties of gradient metasurfaces \cite{Kildishev13,Estakhri16}. However, as the phase gradient originates from resonance coupling between electromagnetic wave and the scatterers, the metasurface performance is essentially limited to near-resonance frequencies.

It was demonstrated that anomalous reflection and refraction of light are governed by generalized Snell's law \cite{Yu11}
\begin{align} \label{Eq:Snell}
&\sin\theta_r - \sin\theta_i = \frac{1}{k_0} \frac{d \Phi(\omega, x)}{d x}, \\
&n_t \sin\theta_t - \sin\theta_i = \frac{1}{k_0} \frac{d \Phi(\omega, x)}{d x}
\end{align}
for the metasurface residing in $x-y$ plane, and with a phase change gradient $\Phi(\omega, x)$ created in $x$ direction only. The medium above the metasurface is vacuum, while the medium below metasurface has refractive index $n_t$, $\theta_i$, $\theta_r$ and $\theta_t$ are angles of incidence, reflection, and refraction, respectively, $k_0 = \omega/c$ is free-space wavenumber, $\omega$ is radiation frequency. One can clearly see from Eq. (\ref{Eq:Snell}), that the necessary condition for achromatic reflection and refraction in a broad frequency range is for a phase gradient to be a separable function of the form 
\begin{equation} \label{Eq:phase}
\Phi(\omega, x) = k_0 X(x),
\end{equation}
as in this case reflection and refraction angles do not depend on frequency, i.e.
\begin{align} \label{Eq:Snell_freq}
&\sin\theta_r - \sin\theta_i = \frac{d X(x)}{d x}, \\
&n_t \sin\theta_t - \sin\theta_i =\frac{d X( x)}{d x}.
\end{align}
The fact that the gradient, $\Phi(\omega, x)$, has to be a linear function of frequency poses a challenge for creating achromatic metasurface using sub-wavelength near-resonance scatterers, as the phase of scattered wave varies abruptly around the resonance frequency, and thus phase gradient defined by Eq. \eqref{Eq:phase} is not easy to implement.

Recently, the problem of broadband achromatic refraction has been addressed in a number of papers \cite{Aieta15,Aieta15a,Li16}. Particularly, achromatic lens operating at three wavelengths (1300 nm, 1550 nm and 1800 nm) has been designed \cite{Aieta15,Aieta15a} utilizing dielectric resonators supporting dense spectrum of optical modes that allows to implement near-resonant coupling at multiple wavelengths. However, the phase change experienced by the wave still varies abruptly around each of the resonances, and thus achromatic behavior can be only implemented for discrete number of frequencies. Alternatively, metal-insulator-metal (MIM) resonators were proposed \cite{Li16} to implement broadband achromatic refraction in near-IR frequency range. In this case incident wave excites gap plasmons in an array of MIM resonators that re-radiate their energy after propagation in the resonator. The phase gradient was implemented by changing resonators width and thus the gap plasmons phase velocities. 

In this paper we address the issue of achromatic anomalous reflection, i.e. we design a mirror that reflects incident light beam at an angle that is defined by the mirror geometry only, and does not depend on radiation wavelength. We utilize aperiodic array of MIM resonators for this purpose and demonstrate that the mirror can operate in near-IR and visible frequency ranges if silver is chosen as a material for resonators design. Particularly, we argue that the metasurface is capable of steering normally incident beam at an angle as high as 40 degrees, while having Joule losses not exceeding 10 $\%$ and radiation efficiency above 98 $\%$. 

\section{Theory}

\begin{figure}
	\includegraphics[width=4in]{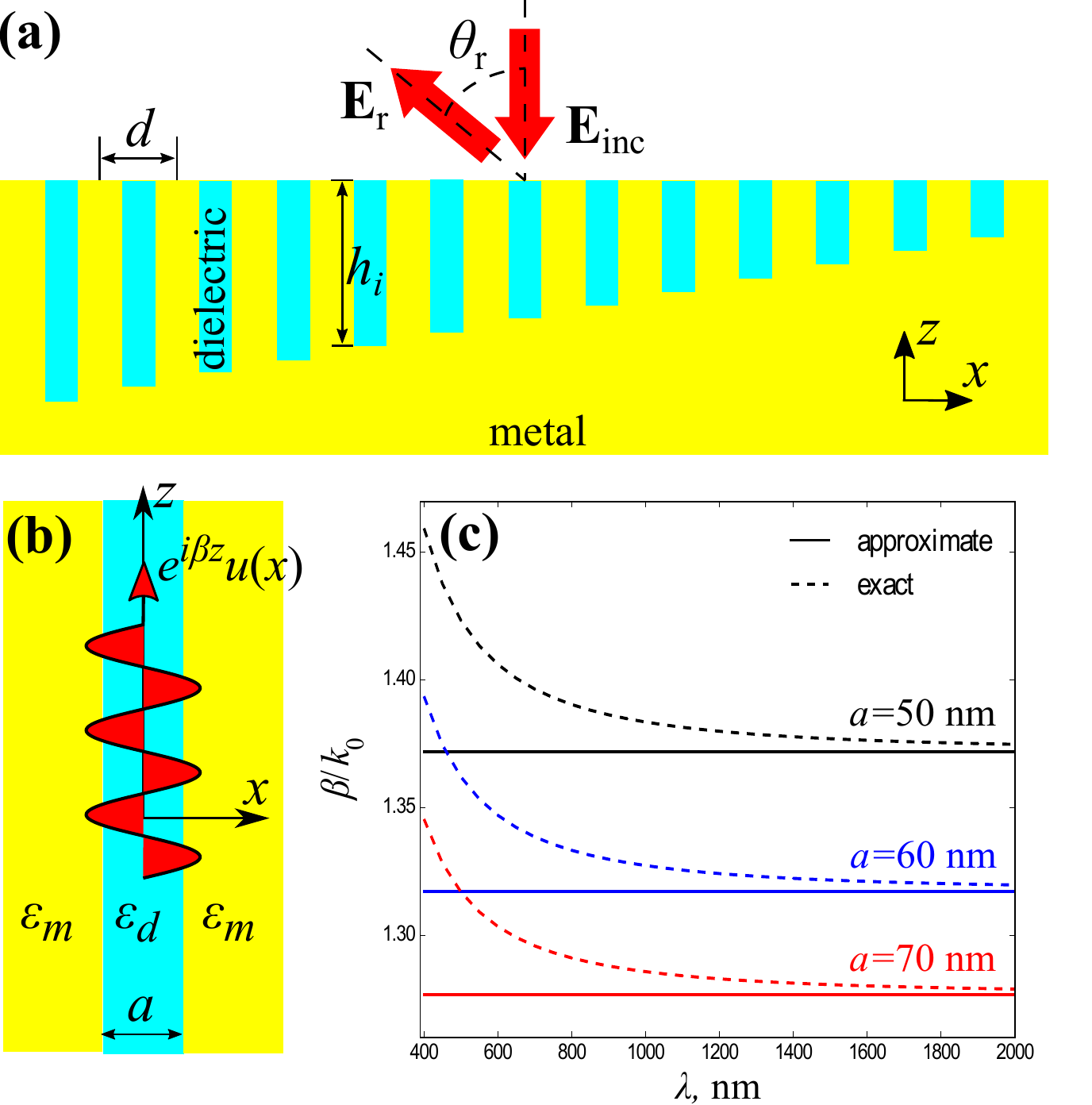}
	\caption{\textbf{(a, b)} Geometry of aperiodic metasurface for steering normally incident beam at angle $\theta_r$ with respect to the surface normal. Metasurface is build from $N$ MIM resonators (panel (b)) of length $h_i$ and has length $D = N d$, where $d$ is a width of individual resonator ($a$ is width of a dielectric in each MIM). Relative permittivities of metal and dielectric are $\varepsilon_m$ and $\varepsilon_d$, respectively. Incident wave, $\mathbf{E}_{inc}$, excites gap plasmons in each of the resonators, with plasmon wavenumber, $\beta$, and electric field distribution of a form $e^{\pm i \beta z} u(x)$. \textbf{(c)} Normalized gap plasmon wavenumber $\beta/k_0$ for different gap widths, $a$, as a function of radiation wavelength for silver-air-silver resonator. }
	\label{Fig1}
\end{figure}

In order to design achromatic anomalous mirror we use metal-insulator-metal (MIM) resonators as building blocks (see Fig. \ref{Fig1}a,b). The wave impinging on such a resonator excites gap plasmon,  $e^{\pm i\beta z} u(x)$, where $\beta$ is a plasmon wavenumber and $u(x)$ is a plasmon electric field distribution in transversal plane (see Fig. \ref{Fig1}b). The additional phase acquired by the plasmon, propagating back and forth inside the resonator before re-radiating its energy back to space, is
\begin{equation}
\Phi  \approx 2 h \mathrm{Re}\beta,
\end{equation}
where $h$ is a resonator length.

The dispersion relation for gap plasmons in MIM structures is well-known and given by\cite{Maier07}
\begin{equation} \label{Eq:disp1}
\tanh \frac{\gamma_d a}{2} = - \frac{\gamma_m \varepsilon_d}{\gamma_d \varepsilon_m},
\end{equation}
where $\varepsilon_{d}$ and $\varepsilon_m$ are relative permittivities of dielectric and metal, respectively, while $\gamma_i = \sqrt{\beta^2 - k_i^2 }$, and $k_i^2 = k_0^2 \varepsilon_i$. In what follows, we assume that in the frequency range under the consideration, the permittivity of metal can be approximated by a Drude model
\begin{equation}
\varepsilon_m = \varepsilon_{\infty} - \frac{\omega_p^2}{\omega^2 + i \gamma\omega},
\end{equation}
where $\varepsilon_{\infty}$ is the metal permittivity at high frequencies, $\omega_p$ is a plasma frequency, and $\gamma$ is a relaxation frequency. In the case of silver these parameters take the values\cite{Ordal85} $\varepsilon_{\infty} = 1$, $\omega_p = 1.37\times10^{16}$ s$^{-1}$, $\gamma = 2.73\times10^{13}$ s$^{-1}$. We are interested in the case when $\gamma_d a \ll 1$ and $\omega_p \gg \omega, \gamma$. In this case it is straightforward to show (see SI) that the plasmon wavenumber takes the form
\begin{align} \label{Eq:disp2}
\beta = k_0 \sqrt{ \varepsilon_d \left(1  + \frac{2c}{a\omega_p}\right)},
\end{align}
and thus phase change experienced by the scattered wave takes the form of Eq. \eqref{Eq:phase}, i.e.,
\begin{equation} \label{Eq:ph_grad}
\Phi(\omega, x) = k_0 2 h(x)\sqrt{ \varepsilon_d(x) \left(1  + \frac{2c}{a(x)\omega_p}\right)},
\end{equation}
where permittivity of the dielectric in the resonator is assumed to be independent of frequency.

Comparison between gap plasmon propagation constants, $\beta$, calculated using exact equation \eqref{Eq:disp1} and approximate equation \eqref{Eq:disp2}, is presented in Fig. \ref{Fig1}c for the case of silver-air-silver resonator and for different gap widths, $a$. One can see that Eq. \eqref{Eq:disp2} is a good approximation for the gap plasmon wavenumber over a broad frequency range from near-IR up to visible. However, the approximation becomes less reliable at higher frequencies beyond $\lambda = 400$ nm ($\omega = 0.47\times 10^{16}$ rad/s) as the radiation frequency becomes comparable to plasma frequency $\omega_p$ in silver. 

\begin{figure}[h!]
\includegraphics[width=5in]{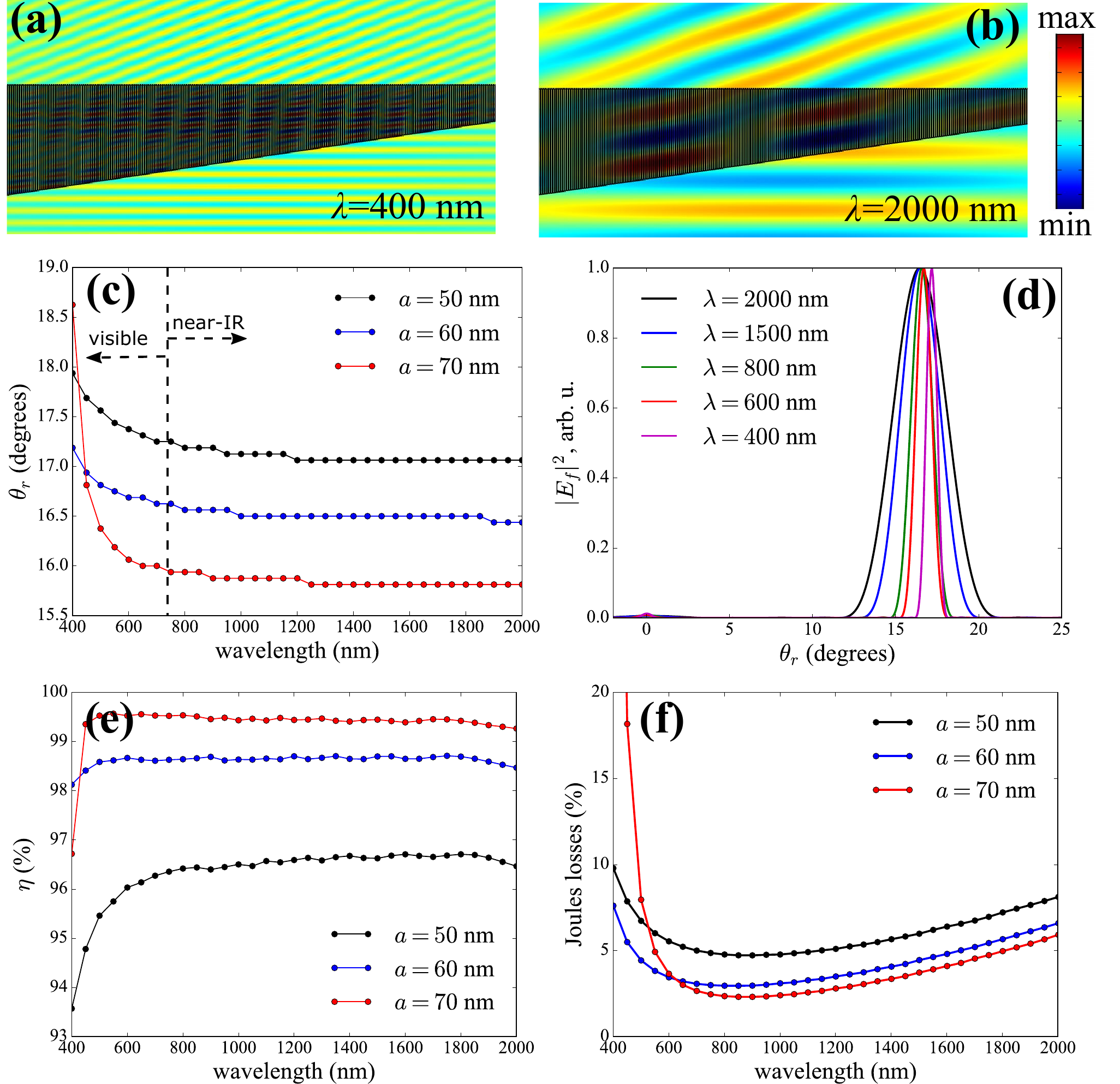}
	\caption{Reflection of normally incident Gaussian beam, defined by Eq. \eqref{Eq:Gaussia}, by a metasurface consisting of $N = 450$ resonators of width $d = 80$ nm each. The total metasurface width is $D = N d = 36 \mu$m. Metasurface is designed to steer beam by angle $\theta_r \approx 17^\textrm{o}$ in the wavelength range from $\lambda = 400$ nm till $2000$ nm. $\alpha = 0.005$ $\mu$m$^{-2}$. \textbf{(a, b)} Spatial distribution of $x$-component of electric field intensity in reflected wave for wavelengths $\lambda = 400$  nm (a) and $\lambda = 2000$ nm (b), assuming that the gap width $a = 60$ nm. \textbf{(c)} Angle of reflection, $\theta_r$, as a function of wavelength for different gap widths $a$. \textbf{(d)} Angular distribution of intensity of reflected wave for different wavelengths. $a = 60$ nm. \textbf{(e)} Radiation efficiency of the metasurface, calculated using Eq. \eqref{Eq:rad} and $\delta\theta = 4^{\circ}$.  \textbf{(f)} Joules losses in the metasurface (see Eq. \eqref{Eq:Joule})  }
	\label{Fig2}
\end{figure}

As one can see from Eq. \eqref{Eq:ph_grad}, there are several different ways to create phase gradient, $\Phi(\omega, x)$. We can use different dielectric materials inside the resonators along the mirror surface (i.e. to vary $\varepsilon_d(x)$). However, this approach is not efficient as the permittivity of the dielectric material should be close to the permittivity of the material above the metasurface in order to suppress specular reflection, which imposes considerable restrictions on the variation of $\varepsilon_d(x)$. Alternatively, we can vary gap widths, $a(x)$. The parameter space, we can explore in this case, is also quite limited as the gap width should be much smaller than the radiation wavelength. Otherwise, we have plasmons propagating along the resonator walls instead of concentrating their electric field in resonator gap, which leads to high losses. Moreover, resonator width can not be too small either in order to prevent specular reflection from metal. In this paper we pursue the third approach, depicted in Fig. \ref{Fig1}a, i.e. we vary resonators lengths, $h$.

\section{Results and discussion}
All the results presented in this section were obtained assuming that the minimum length of resonators in the mirror is $h_{min} = 200$ nm, while the maximum length depends on the number of resonators $N$. Simulations were made using COMSOL 5.2. We use normally incident Gaussian beam,
\begin{align} \label{Eq:Gaussia}
 \mathbf{E}_i(x,z) = \mathbf{e}_x E_0 e^{- i k z} e^{ - \alpha x^2},
 \end{align} 
to illuminate the mirror. The electric field in the beam has to be polarized along the resonator width (i.e. $x$-axis, see Fig. \ref{Fig1}a) in order to excite the gap plasmon, $k = k_0 n$, and parameter $\alpha$ defines spatial width of the wave, i.e. the beam spot size. 

The simulation results are presented in Fig. \ref{Fig2} for a metasurface designed to steer beam by an angle $\theta_r \approx 17^{\textrm{o}}$. We assume normally incident gaussian beam with $\alpha = 0.005$ $\mu$m$^{-2}$ (corresponding to spot size of about 30 $\mu$m). The mirror consists of $N = 450$ silver-air-silver resonators, each $d = 80$ nm wide, and has total width $D = N d = 36$ $\mu$m. We consider three different gap widths, $a = 50$, $60$, and $70$ nm. The spatial distribution of $x$-component of scattered electric field is presented in Figs. \ref{Fig2}a,b. One can clearly see that wavefront of the reflected wave propagates at an angle with respect to the normal to the metasurface\footnote{Note that scattered electric field, $\mathbf{E}_{sc}$, in Figs. \ref{Fig2}a,b is non-zero inside the metal. This is because it is total electric field, $\mathbf{E}_{tot} = \mathbf{E}_{i} + \mathbf{E}_{sc}$, that has to be zero inside the metal and not the scattered electric field. The spatial distribution of total electric field is presented in SI}. The angle of anomalous reflection is presented in Fig. \ref{Fig2}c as determined from the angular distribution of the intensity carried by the reflected wave, shown in Fig. \ref{Fig2}d. Here we took into account that Poynting vector of a plane wave is proportional to $|\mathbf{E}|^2$. The results presented in Fig. \ref{Fig2} clearly demonstrate that the metasurface is achromatic for the wavelengths between 400 nm and 2000 nm. Indeed, for all three gap widths, difference in steering angles at 2000 nm and 500 nm is less than 0.5$^{\circ}$ (see Fig. \ref{Fig2}c). This difference increases around 400 nm (as the approximation defined by Eq. \eqref{Eq:disp2} becomes less reliable, see Fig. \ref{Fig1}c), but even in this case the difference does not exceed 1$^{\circ}$ for gap widths $a = 50$ and $60$ nm. When the gap width becomes sufficiently large ($a=70$ nm), however, the difference in reflection angles at 2000 nm and 400 nm becomes as large as 3$^{\circ}$. The non-zero angular width of the reflected beam, that one can see on Fig. \ref{Fig2}d, is due to the spatial localization of the beam. One can see that the angular spread decreases with the decrease of the wavelength. Indeed, the direction of the energy flow in plane wave is defined by wavevector $\mathbf{k}$. The Fourier decomposition of Gaussian beam defined by Eq. \eqref{Eq:Gaussia}, apart from the main component $k \mathbf{e}_z$, contains additional components, $k \mathbf{e}_z + \Delta k \mathbf{e}_x$, that carry energy at some angle with respect to the normal to the mirror plane. The angular spread of the incident beam $\Delta k$ is proportional to $\lambda/\Delta x$ , i.e. to the ratio between the radiation wavelength and spot size of the beam, $\Delta x \propto 1/\alpha$. The reflection of these obliquely incident components causes angular spread of the reflected beam.  

In order to characterize performance of the mirror, we calculated absorption losses and radiation efficiency. We define radiation efficiency as the ratio between the intensity of the radiation scattered by the metasurface in a given angle to the total intensity of the scattered radiation, i.e.
\begin{equation} \label{Eq:rad}
\eta = \frac{\int\limits_{\theta_r - \delta\theta}^{\theta_r + \delta \theta} |E_f(\theta)|^2 d\theta}{\int\limits_{0^{\circ}}^{360^{\circ}} |E_f(\theta)|^2 d\theta},
\end{equation}
where we took into account that Poynting vector of a plane wave, and thus the intensity of radiation in a given angle, are proportional to $|E_f(\theta)|^2$, where $E_f(\theta)$ is the angular distribution of the electric field intensity in the far-field. Angle $\delta\theta$ accounts for the angular spread of the reflected beam discussed above. Thus the radiation efficiency accounts for the losses due to the specular reflection and wave scattering into higher diffraction orders. However, we also need to account for the fraction of the incident radiation absorbed by the metal in the mirror and thus converted into heat. We account for this by defining Joules losses as the ratio between power $Q$ converted into heat and total power delivered to the metasurface by the incident beam, i.e.
\begin{equation} \label{Eq:Joule}
J = \frac{Q}{P_{in}},
\end{equation}
where $Q$ is calculated numerically using COMSOL, while incident power is estimated as $P_{i} = (1/2\eta) \int_S |\mathbf{E}_i|^2 dS $, where $\eta$ is the wave impedance in the medium above the metasurface, and $S$ is the surface of the metasurface. 

Calculated radiation efficiency and Joules losses are presented in Fig. \ref{Fig2}e,f. One can see (Fig. \ref{Fig2}e) that radiation efficiency of the metasurface can be as high as 99$\%$ in a broad frequency range, when gap width, $a = 70$ nm, is close to the resonator width, $d = 80$ nm. The radiation efficiency, however, gradually decreases with decrease of gap width ($\eta \approx 96 \%$ for $a = 50$ nm) due to increase of specular reflection from the metal. Absorption losses (see Fig. \ref{Fig2}f), on the other hand, increase with the decrease of gap width. This can be attributed to increase in plasmon losses as, in the case of narrower gaps, electric field of the plasmon penetrates deeper into the metal. The deeper penetration of electromagnetic field into the silver is also responsible for increase of absorption losses for wavelengths above 800 nm (i.e. for small values of $a/\lambda$). The absorption losses increase for shorter wavelengths as silver itself gets more absorptive and less reflective in the optical range. Thus, we conclude that in the general case it is beneficial to use resonators with large gap width, as they tend to have lower absorption losses and higher radiation efficiencies. In particular, for a metasurface operating in the range between 500 nm and 2000 nm, gap width $a = 70$ nm is about optimal. However, in order to extend the metasurface functionality up to 400 nm, resonators with a smaller gap width of $a = 60$ nm provide a better trade-off, as they can alleviate high absorption losses.   

\begin{figure}[h!]
	\includegraphics[width=\textwidth]{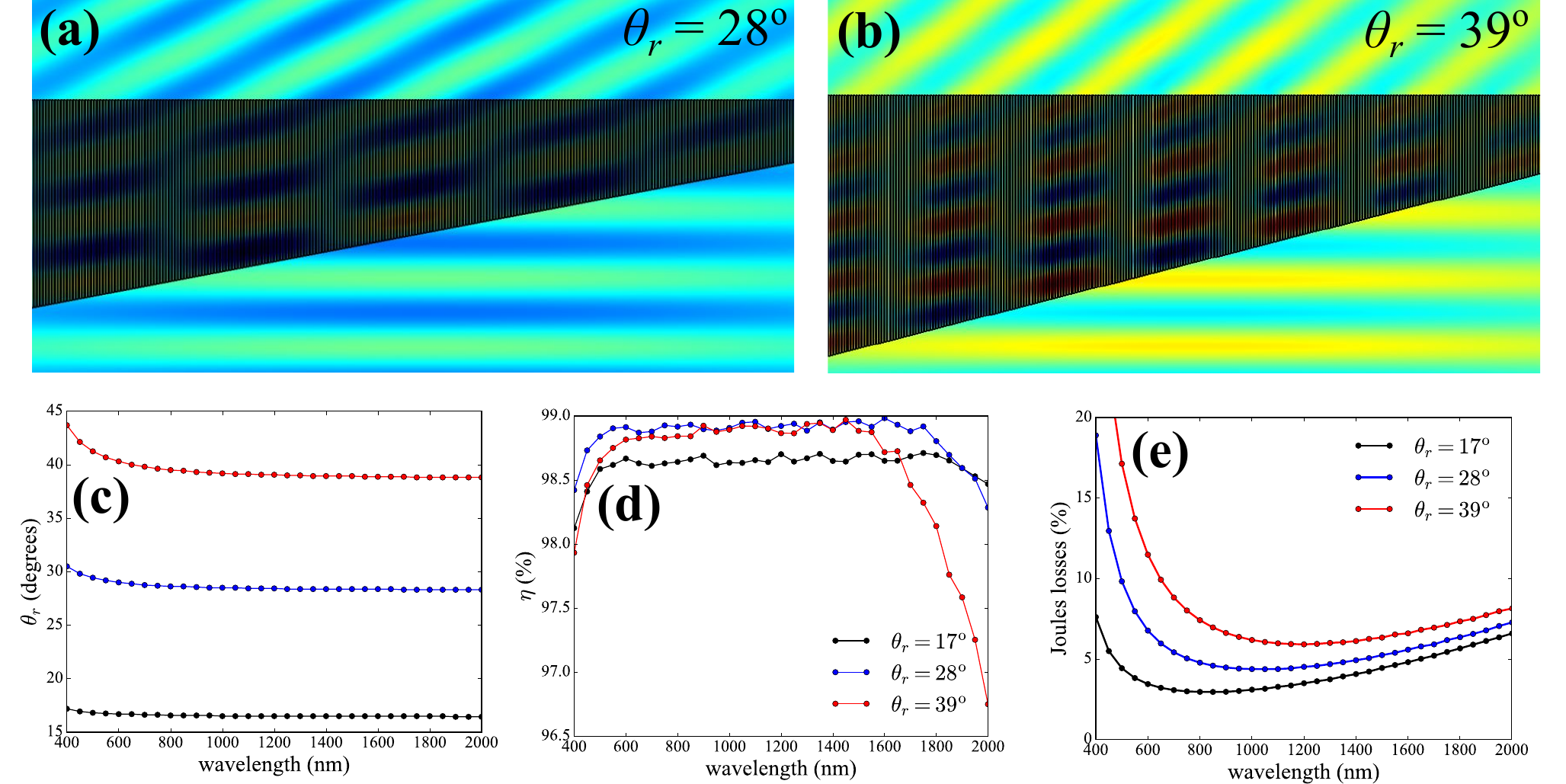}
	\caption{Reflection of normally incident Gaussian beam, defined by Eq. \eqref{Eq:Gaussia}, by a metasurface consisting of $N = 450$ resonators of width $d = 80$ nm each. Gap width $a = 60$ nm. $\alpha = 0.005$ $\mu$m$^{-2}$. \textbf{(a, b)} Spatial distribution of $x$-component of electric field intensity in reflected wave at the wavelength $\lambda = 2000$  nm for angles of reflection $\theta_r = 28^{\circ}$ (a) and $\theta_r = 39^{\circ}$ (b). \textbf{(c,d,e)} Wavelength dependence of the reflections angle $\theta_r$ (c), radiation efficiency (d), and absorption losses (e).}
	\label{Fig3}
\end{figure}

As a next step we study potential of the anomalous mirror for steering the normally incident beam to angles larger than 17$^{\circ}$. This requires creating larger phase gradient along the metasurface, which can be implemented by increasing difference between maximum, $h_{max}$, and minimum, $h_{min}$, lengths of the MIM resonators, while keeping number of resonators, $N$, and their widths, $d$, the same. The simulation results are presented in Figure \ref{Fig3} for three different steering angles, $\theta_r = 17^{\circ}$ ($h_{max} = 4.6 \mu$m), $\theta_r = 28^{\circ}$ ($h_{max} = 7.6 \mu$m), and $\theta_r = 39^{\circ}$ ($h_{max} = 10 \mu$m). The minimum length of the resonators was the same in all three cases, $h_{min} = 0.2 \mu$m. One can see from Fig. \ref{Fig3} that for wavelengths between 600 nm and 2000 nm the mirror is achromatic, has radiation efficiency as high as 95$\%$ and Joules losses smaller than $10\%$ even if reflection angle is as high as $40^{\circ}$. At the shorter wavelengths reflection angle deviates significantly from that at 2000 nm (around 5$^{\circ}$ for $\theta_r = 39^{\circ}$), moreover absorption losses increase drastically (more than 20$\%$ around 400-500 nm for $\theta_r = 39^{\circ}$). In general, the absorption losses tend to increase with the increase of the reflection angle as the achromatic mirror for large angles is built from longer resonators. This leads to longer travel distances for gap plasmons, and correspondingly to higher attenuation. We can, however, conclude that the proposed design for the mirror can be used for efficient achromatic light steering at angles as large as 40$^{\circ}$ in the near-IR frequency ranges. 

The design of achromatic mirror proposed so far relied on utilizing aperiodic metasurface for achromatic beam steering. The natural question, however, if it is possible to create periodic achromatic mirror utilizing aperiodic chunks as building blocks. Indeed, in the case of mirror that steers beam at  $17^{\circ}$ angle (see Fig. \ref{Fig2}), the smallest resonator had length $0.2\mu$ m and a wave impinging on this resonator experiences phase change of about $2 \mathrm{Re}\beta h_{min} \approx 0.53\pi$ (at the wavelength 2000 nm). On the other hand, the wave impinging on the longest resonator of length $4.6\mu$m experiences phase change of about $12.14 \pi$. Thus the total phase change along the mirror significantly exceeds $2\pi$ and it seems that periodic mirror can be created from smaller aperiodic chunks providing $2\pi$ phase difference. The problem, however, arises from the fact that the phase change experienced by the wave interacting with a single resonator is wavelength dependent (see Eq. \eqref{Eq:ph_grad}). In order to clarify this point, let us assume that we have an aperiodic mirror (consisting of $N$ resonators of width $d$) that provides phase change equals to $2\pi$ at a wavelength $\lambda_1$. This condition obviously imposes restriction on the maximum and minimum lengths of the resonators in the metasurface such as $k_1 2 (h_{max} - h_{min}) l = 2\pi$, where $l = \sqrt{ \varepsilon_d \left(1  + 2c/a\omega_p\right)}$. It is obvious, that for a given geometry the above condition can be satisfied only at a single wavelength. However, we can reformulate this condition in more general form $k_n 2 (h_{max} - h_{min}) l = 2\pi n$, or
\begin{equation} \label{Eq:wave}
\lambda_n = 2 (h_{max} - h_{min}) l/n.
\end{equation}
Thus periodic metasurface can operate at multiple discrete wavelengths. 

\begin{figure}[h!]
	\includegraphics[width=5in]{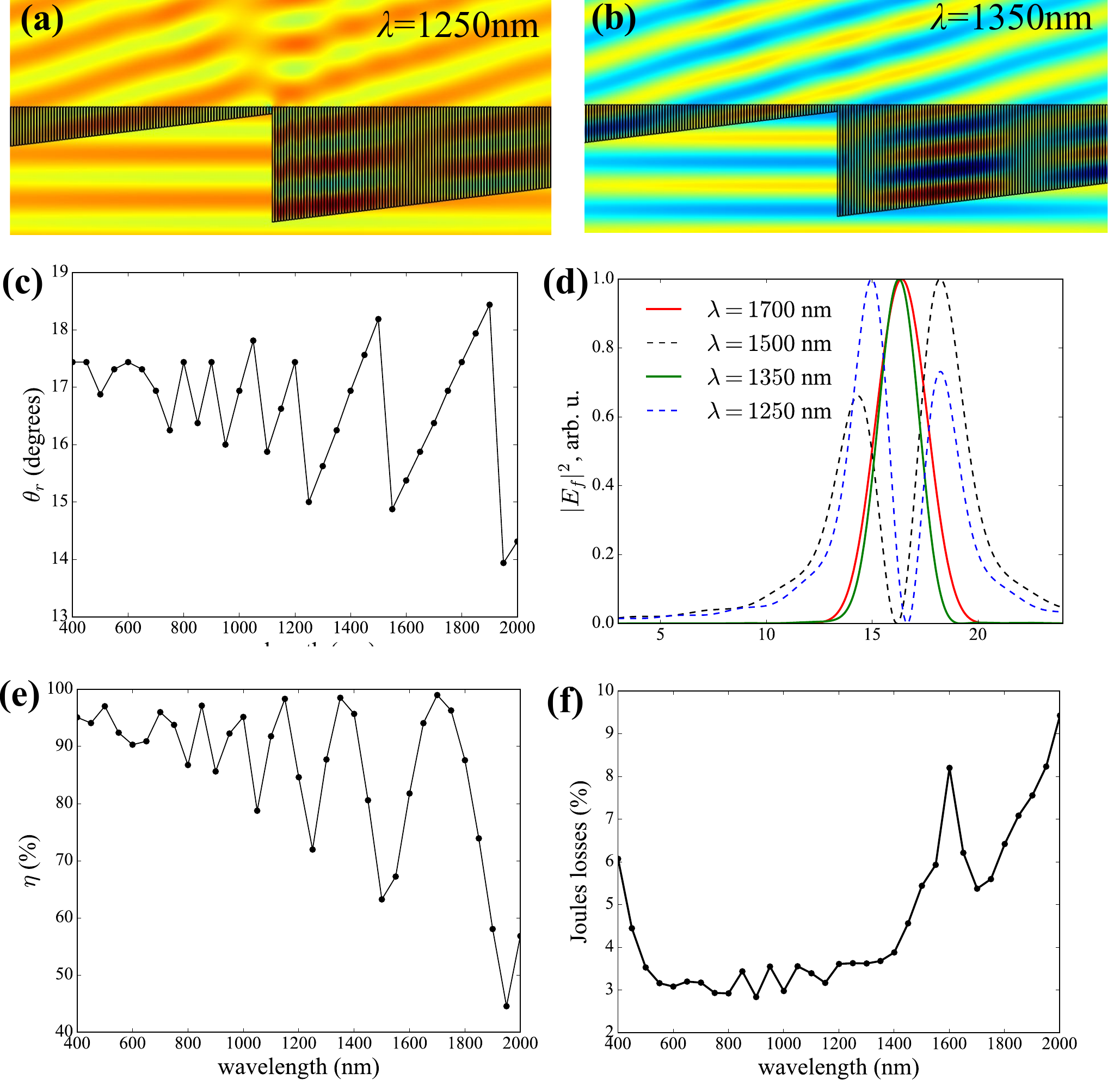}
	\caption{Reflection of normally incident Gaussian beam, defined by Eq. \eqref{Eq:Gaussia}, by a metasurface consisting of two aperiodic arrays of resonators containing $N = 300$ resonators each. Resonators width $d = 80$ nm each. Gap width $a = 60$ nm. $\alpha = 0.005$ $\mu$m$^{-2}$. Each of two arrays is designed to steer beam by angle $\theta_r \approx 17^\textrm{o}$. \textbf{(a, b)} Spatial distribution of $x$-component of electric field intensity of reflected wave at wavelengths $\lambda = 1250$  nm (a) and $\lambda = 1350$  nm (b). \textbf{(c)} Angle of reflection, $\theta_r$, as a function of wavelength. \textbf{(d)} Angular distribution of intensity of reflected wave for different wavelengths. \textbf{(e)} Radiation efficiency of the metasurface, calculated using Eq. \eqref{Eq:rad} and $\delta\theta = 4^{\circ}$.  \textbf{(f)} Joules losses in the metasurface (see Eq. \eqref{Eq:Joule}).}
	\label{Fig4}
\end{figure}

In order to illustrate this point in more details we did simulations for metasurface built from two aperiodic identical arrays of resonators placed side by side (see Fig. \ref{Fig4}). Each of the arrays contained $N = 300$ resonators and was designed to steer beam at angle $\theta_r = 17^{\circ}$. Simulations results are presented in Fig. \ref{Fig4}. One can clearly see distortion in reflected wavefront at some wavelengths ($\lambda = 1250$ nm, Fig. \ref{Fig4}a). This distortion is due to the fact that at these wavelength total phase difference across the aperiodic array of 300 resonators is not multiple of $2\pi$. On the other hand there is no distortion of the wavefront at $\lambda = 1350$ nm, where the phase difference is multiple of $2\pi$. The distortion of the wavefront leads to the splitting of a reflected beam (see Fig. \ref{Fig4}d, $\lambda = 1250$ and 1500 nm) into two slightly detuned angular directions (Fig. \ref{Fig4}c). This in turn leads to strong oscillations of the radiation efficiency of metasurface. For example, at $\lambda = 1500$ nm the efficiency drops to $60\%$ as the $40\%$ of the radiation is carried by a second beam (see Fig. \ref{Fig4}b). Nevertheless, as was predicted by Eq. \eqref{Eq:wave}, there is a discrete set of frequencies for which metasurface produces a single beam ($\lambda = 1700$, 1350, 1150 and 1000 nm) and radiation efficiency of the metasurface exceeds $95\%$ at these frequencies.

\section{Conclusions}
Concluding, we presented the design of achromatic anomalous mirror operating in the near-IR and visible frequency ranges. We used an array of metal-insulator-metal (MIM) resonators as building blocks of the mirror. An electromagnetic wave impinging on MIM resonator launches plasmons propagating inside the resonator gap and phase change experienced by the wave is proportional to the plasmon optical path. The phase gradient can then be created along the mirror by using the array of resonators of different lengths. When the frequency of the electromagnetic wave is much smaller than the plasmon frequency in the metal, plasmon propagation constant, and thus the phase gradient, is a linear function of frequency, which is a sufficient condition for an achromatic anomalous reflection. 
We demonstrated that mirror comprised from silver-air-silver resonators can sustain achromatic regime in the visible and near-IR frequency ranges. In the case of normally incident Gaussian beam, the Joule losses in such a mirror do not exceed 10 $\%$, while radiation efficiency exceeds 98 $\%$ for steering angles as high as 40$^{\circ}$. 

\section*{acknowledgement}
This work is supported by a DARPA grant award FA8650-16-2-7640. We acknowledge useful discussions with Predrag Milojkovic, Jay Lewis and Ramzi Zahreddine.

\end{document}